# Direct evidence of long-range ordering of reduced magnetic moments in the spin-gap compound CeOs$_2$Al$_{10}$ through muon spin relaxation investigations

D.T. Adroja[1], A.D. Hillier[1], P.P. Deen[2], A.M. Strydom[3], Y. Muro[4], J. Kajino[4], W.A. Kockelmann[1], T. Takabatake[4], V.K. Anand[1], J.R. Stewart[1] and J. Taylor[1]

[1]ISIS Facility, Rutherford Appleton Laboratory, Chilton, Didcot Oxon, OX11 0QX, UK
[2]Institute Laue- Langevin, BP 156, 6 Rue Jules Horowitz 38042, Grenoble Cedex, France
[3]Physics Department, University of Johannesburg, PO Box 524, Auckland Park 2006, South Africa
[4]Department of Quantum matter, ADSM, and IAMR, Hiroshima University, Higashi-Hiroshima, 739-8530, Japan

(Dated: 25$^{th}$ July 2010)

We have carried out neutron diffraction, muon spin relaxation (μSR) and inelastic neutron scattering investigations on a polycrystalline sample of CeOs$_2$Al$_{10}$ to investigate the nature of the phase transition observed near 29 K in the resistivity and heat capacity. Our μSR data clearly reveal coherent frequency oscillations below 28 K, indicating the presence of an internal field at the muon site, which confirms the long-range magnetic ordering of the Ce-moment below 28 K. Upon cooling the sample below 15 K, unusual behaviour of the temperature dependent μSR frequencies may indicate either a change of the muon site, consistent with the observation of superstructure reflections in electron diffraction or a change of the ordered magnetic structure. Neutron diffraction data do not reveal any clear sign of either magnetic Bragg peaks or superlattice reflections. Furthermore, inelastic neutron scattering (INS) measurements clearly reveal the presence of a sharp inelastic excitation near

11 meV between 5 K and 26 K, due to opening of a gap in the spin excitation spectrum, which transforms into a broad response at and above 30 K. The magnitude of the spin-gap (11 meV) as derived from the INS peak position agrees very well with the gap value as estimated from the bulk properties.

# I. Introduction

The Ce-based strongly correlated electron systems have attracted considerable interest due to the duality between the itinerant and the localized nature of *4f*-electrons that gives rise to a rich variety of novel phenomena, such as heavy electron behaviour, mixed valence behaviour, reduced magnetic moment ordering, unconventional superconductivity, Kondo insulator or Kondo semiconductors, spin and charge gap formation, charge and spin density waves, metal-insulator transition, spin-dimer formation, non-Fermi-liquid (NFL) behavior and quantum criticality [1-6]. The variety of phenomena arise due to the presence of strong electron-electron correlations as revealed by a large enhancement of low temperature properties such as the electronic heat capacity coefficient and static susceptibility above the values expected from local density approximation electronic-structure calculations [7]. The consequences of a variable degree of hybridization between 4f and conduction electrons on electronic and magnetic properties in these systems have given genesis to a wealth of novel features that have kept this materials class at the forefront of condensed matter physics for more than three decades. Whereas the single-ion Kondo picture has been a major achievement in our understanding of moment formation in solids, the many-body cooperative phenomena arising from a regular lattice of local moments, or Kondo lattice, still evades a comprehensive description.

Recent reports on magnetic and transport properties of Ce-based ternary compounds of type $CeT_2Al_{10}$ (T=transition metals), which crystallise in the orthorhombic structure (space group No 63 Cmcm), have generated considerable interest in both theoretical and experimental condensed matter physics [8-15]. This interest arose due to the various interesting ground states observed in this family of Ce-compounds. An unusually sharp phase transition near 27 K in the magnetic susceptibility of $CeRu_2Al_{10}$ [10] has been attributed to a spin-dimer formation [14, 15]. The resistivity of $CeRu_2Al_{10}$ exhibits a sharp drop near 27 K resembling an insulator-metal transition [10-12]. A very similar phase transition, near 29 K, has been observed in $CeOs_2Al_{10}$ [11,12], but in this compound the susceptibility (along a-axis) exhibits a broad maximum near 45 K in contrast to a sharp drop at the phase transition (27 K) in $CeRu_2Al_{10}$ [10]. The broad maximum in the susceptibility and its strong anisotropic behavior in the paramagnetic state reveal the presence of strong hybridization between 4f- and conduction electrons as well as strong single ion anisotropy arising from the crystal field potential. Another difference between the two systems appears in the resistivity of the ordered state; the resistivity of $CeOs_2Al_{10}$ displays a thermal activation-type temperature dependence below 15 K while the resistivity of $CeRu_2Al_{10}$ exhibits metallic behaviour below the phase transition down to 2 K.

Furthermore, the observed phase transitions in these two compounds have some resemblance to that of the hidden order transition observed in $URu_2Si_2$ with very small ordered state moment at 17 K [16]. Therefore, systematic investigations of $CeT_2Al_{10}$ (T = Fe, Ru, and Os) with different values of the Kondo temperature $T_K$ are necessary to reveal the role of the c-f hybridization in the mysterious phase transition and gap formation. In spite of the comparable transition temperatures between $CeRu_2Al_{10}$ (27 K) and $CeOs_2Al_{10}$ (29 K), a fundamental contrast in the electronic disposition of the two compounds has been exposed in very recent high-pressure studies [11]. The electrical resistivity of $CeRu_2Al_{10}$ in an applied hydrostatic pressure of 1.75 GPa has been shown to match the overall behaviour of temperature dependent resistivity of $CeOs_2Al_{10}$ in zero applied pressure very closely. Applying pressure is well recognised as a tuning method of the 4f-band with respect to the Fermi energy ($E_F$) in narrow-band systems. These observations among the two iso-electronic compounds are indication that the center of gravity of the 4f-band is lying on opposite sides of $E_F$, where the 4f spectral weight in $CeRu_2Al_{10}$ is most likely close to but below $E_F$ whereas in $CeOs_2Al_{10}$, $E_F$ is most likely underlying the 4f band. This is in agreement with the view of a more extended nature of the 5d band of Os compared to the 4d band of Ru. Furthermore, $CeFe_2Al_{10}$ exhibits Kondo insulating behaviour with a transport gap of 15 K, while an NMR study reveals a larger value of the gap, namely 110 K [17]. The Kondo insulator behaviour observed in $CeFe_2Al_{10}$ has some similarity with that of the well-known Kondo insulator compounds CeNiSn and CeRhSb [18, 19].

Very recently neutron scattering and µSR studies on $CeRu_2Al_{10}$ have been performed by Robert et al [20], our group [21], and Kambe et al [22]. The inelastic neutron scattering study (INS) reveals a clear sign of the spin-gap formation of 8 meV [19-20]. The gap is nearly temperature independent up to 24 K, but then suddenly disappears at 27 K. By raising the temperature still further, the INS response becomes very broad, of quasi-elastic-type [20-21]. The behavior of both the gap magnitude and its temperature dependence are in good agreement with predictions based on a theoretical model for a spin-dimer formation pertinent to this class of compounds which has recently been forwarded by Hanzawa [15]. A µSR study to probe the nature of magnetic ordering in $CeRu_2Al_{10}$ revealed the presence of a small internal field (120 G [21] and 20 G [22]) at the muon stopping site in zero-field indicating the long-range magnetic ordering of the $Ce^{3+}$ moment [21,22]. It is interesting to note that the µSR frequency, which is a measure of internal magnetic field seen by muons, decreases as the temperature falls below 20 K suggesting a further phase transition at low temperature [22].

We present in this paper our results of µSR, neutron diffraction and inelastic neutron scattering studies on $CeOs_2Al_{10}$ to shed light on the nature of the phase transition and the ground state of the Ce ion in this compound. Prior evidence having alluded to small moment ordering in $CeRu_2Al_{10}$, the method of µSR is an exceptionally sensitive microscopic probe of cooperative magnetic ordering phenomena and is thus ideally suited to this problem. Inelastic neutron scattering gives direct information about the magnitude of the spin-gap energy, its temperature and wave-vector (Q) dependency, which are important to understand the nature of mechanism of the spin gap formation, i.e. whether or not the spin gap originates due to a single ion type mechanism. In this case one expects a Q-independent spin gap to follow the $Ce^{3+}$ form factor squared, $F^2(Q)$. In case of inter-site (or lattice effects) one expects a modulation of the gap value and its intensity with Q.

## II. Experimental details

The polycrystalline sample of $CeOs_2Al_{10}$ was prepared by argon arc melting of the stoichiometric constituents with the starting elements , Ce 99.9% (purity), Os 99.9% and Al 99.9999%. The sample was annealed at 1000 $^o$C for four days in an evacuated quartz ampoule. The sample was characterised using power X-ray diffraction and neutron diffraction and was found to be dominantly single-phase. The impurity phase was found to be about 3%, but its real composition is not known at present.

The µSR experiments were carried out using the MuSR spectrometer in longitudinal geometry at the ISIS pulsed neutron and muon source, UK. At the ISIS facility, a pulse of muons is produced every 20 ms and has a FWHM of ~70 ns. These muons are implanted into the sample and decay with a half-life of 2.2 µs into a positron which is emitted preferentially in the direction of the muon spin axis. These positrons are detected and time stamped in the detectors which are positioned before, F, and after, B, the sample. The positron counts, $N_{F,B}(t)$, have the functional form

$$N_{F,B}(t)=N_{F,B}(0)\exp(-t/\tau_\mu)(1\pm G_z(t)) \qquad (1)$$

where $G_z(t)$ is the longitudinal relaxation function. $G_z(t)$ is determined using

$$G_z(t)=(N_F(t)-\alpha N_B(t))/(N_F(t)+ \alpha N_B(t)) \qquad (2)$$

where $\alpha$ is a calibration constant which was determined at ~200 K by applying a small transverse field (~20 G) and adjusting its value until the resulting damped cosine signal was oscillating around zero. This calibration constant takes into account detector efficiency and the absorption of positrons in the sample and surrounding equipment. The powdered sample (thickness ~1.5mm) was mounted onto a 99.995+% pure silver plate using GE-varnish and was covered with 18 micron silver foil. The neutron diffraction measurements were performed using the GEM diffractometer at ISIS which has six detector banks covering angles from 9.7 to 164 degrees, and crystallographic d-spacings from 0.3 to 40 Å. The powder sample of $CeOs_2Al_{10}$ (12.5 g) was mounted in a 10 mm diameter vanadium can, which was cooled down to 2 K using a standard He-4 cryostat. The inelastic neutron scattering measurements on $CeOs_2Al_{10}$ (17 g sample) were carried out using the MARI time-of-flight chopper spectrometer at ISIS. The sample was wrapped in a thin Al-foil and mounted inside a thin-walled cylindrical Al-can, which was cooled down to 4.5 K inside a top-loading closed-cycle-refrigerator (TCCR) with He-exchange gas around the sample. The measurements were performed with an incident neutron energy $E_i$ = 25 meV, with an elastic resolution (at zero energy transfer) of 0.61 meV (FWHM) and inelastic resolution of 0.43 meV at 11 meV energy transfer.

## III. Results and discussions

### (1) Neutron diffraction

Figure 1 shows the neutron diffraction patterns, collected from one of the low-angle GEM detector banks at 18 degree at 2 K, 20 K, and 35 K. Comparing the data collected at these three temperatures, we do not observe any sign of additional reflections at 20 K or 2 K compared to the 35 K diffraction patterns. The absence of magnetic Bragg peaks at 20 K and 2 K suggests that the ordered magnetic moment of the $Ce^{3+}$ ion must be smaller than about 0.20(±0.05) $\mu_B$. It is to be noted that the value of the effective paramagnetic moment ($\mu_{eff}$) of the Ce ion estimated from the susceptibility is 2.7$\mu_B$, but the high field magnetization measurements give the moment value of 0.11$\mu_B$ at 0.3K and at 9.5T [23]. This is in agreement with the observation of very small internal magnetic fields seen in the µSR data as will be discussed in the next section. It is to be noted that neutron diffraction data [21, 24] and also the report by J. Robert et al. [20] on $CeRu_2Al_{10}$ reveal the presence of very weak additional Bragg peaks at 1.8 K. On the basis of (hkl)-indices of the extra peaks at 1.8 K in $CeRu_2Al_{10}$, which are not allowed in the Cmcm space group, J. Robert et al proposed the change of space group symmetry from Cmcm to Amm2 or Pmmn below the phase transition [20]. Moreover, a very recent electron diffraction (ED) study on $CeOs_2Al_{10}$ revealed the presence of new superlattice reflections characterized by a wave vector of q=(0 -2/3 2/3), indicating tripling of the unit cell along [0 -1 1] directions [23]. Our search for the new reflections ascribable to a tripling of lattice parameters along the b and c directions in $CeOs_2Al_{10}$ in the GEM data at 2 K turned out to be fruitless, which may suggest zero or weak structure factors for the superlattice reflections.

Very recently Tanida et al [14] and Hanzawa [15] have proposed a model for $CeRu_2Al_{10}$, for which the phase transition near 27 K is attributed to a spin-Peierls (spin-dimer) transition. In this case one would expect a change of crystal symmetry and also the possible reduction in the Ce-Ce distance due to the formation of the spin-dimer. In order to check the existence of any abrupt change in the lattice parameters and Ce-Ce distances below the phase transition, $T_N$=29 K in $CeOs_2Al_{10}$, we have carried out a full structural refinement of the neutron diffraction data using the GSAS program. At all temperatures, the refinement confirms that the $CeOs_2Al_{10}$ compound crystallizes in the orthorhombic $YbFe_2Al_{10}$-type structure (space group Cmcm, No. 63). The Ce atom is surrounded by a polyhedron formed by 4 Os and 16 Al atoms. The nearest-neighbour Ce-Ce atoms ($d_{Ce-Ce}$ = 5.266(2) Å at 300 K) construct a zigzag chain along the orthorhombic c-axis. The refined lattice parameters, atomic position parameters, thermal parameters and the selected Ce-Ce interatomic distances of $CeOs_2Al_{10}$ are given in Table-I. One can see from Table-I that the lattice parameters (a, b, c) decrease gradually with decreasing temperature. Correspondingly, the unit cell volume is found to decrease linearly going from 35 K to 2 K. Furthermore, the nearest neighbour Ce-Ce distance remains almost unchanged between 2 K and 35 K: $d_{Ce-Ce}$=5.260 (2) Å at 2 K, 20 K and 35 K. Moreover, we have not observed any noticeable changes of Ce-Ce and Ce-Al bond distances and bond angles between 2 K and 35 K. The thermal parameters of all the atoms (see Table 1) are as expected, i.e. no increased values detected for the Ce (or other) atoms, which one would expect for a caged structure. It is interesting to mention that in skutterudite compounds, where the rare earth atoms occupy a caged position in the crystal structure, the values of thermal parameters of the rare earth atoms are about a factor 5-10 larger than those of non-caged atoms [25]. The question of rattling modes in the present 1:2:10 series of compounds remains equivocal. Evidence of optical phonon modes have been found in a systematic study of the related $RFe_2Al_{10}$ (R=Pr and Y) compounds by heat capacity measurements [26].

## (2) μSR measurements

Figure 2 shows the zero-field (ZF) μSR spectra at various temperatures of $CeOs_2Al_{10}$. The insets show the behavior for the shorter time scale. It is interesting to see a dramatic change in the time-evolution of the μSR spectra with temperature. At 35 K we observe a strong damping at shorter time, and the recovery at longer times, which is a typical muon response to nuclear moment, described by the Kubo-Toyabe formalism [27], arising from a static distribution of the nuclear dipole moment. Here it arises from the Os (stable isotope $^{189}Os$, I=3/2, 16.2% abundance, $^{187}Os$, I=1/2, 1.6% abundance) and Al (I=5/2) nuclear moment contributions (I=0 for Ce, i.e. zero contribution). Above the anomaly at 29 K, i.e in the paramagnetic state, the μSR spectra can all be described by the following equation (see Fig. 2, top):

$$G_z(t) = A_0 \left( \frac{1}{3} + \frac{2}{3} (1 - (\sigma t)^2) \exp\left(-\frac{(\sigma t)^2}{2}\right) \right) \exp(-\lambda t) + C \qquad (3)$$

where $A_0$ is the initial asymmetry, $\sigma$ is nuclear depolarization rate, $\sigma/\gamma_\mu = \Delta$ is the local Gaussian field distribution width, $\gamma_\mu$=13.55 MHz/T is the gyromagnetic ratio of the muon, $\lambda$ is the electronic relaxation rate and C is the constant background. It is assumed that the electronic moments give an entirely independent muon spin relaxation channel in real time. The value of $\sigma$ was found to be 0.32 μs$^{-1}$ from fitting the spectra of 29 K to Eq.(3) and was found to be temperature independent between 29 K and 65 K. Using this value together with the crystal structure determined from the neutron diffraction data, we have carried out a finite element analysis, a detailed description of which can be found in ref [29].

Figure 3a shows the unit cell with possible muon positions and Fig. 3b shows an iso-surface plot of the simulated muon nuclear depolarization rate ($\sigma$) arising from the static nuclear moments. The red surface contour (for online color) in the unit cell (Fig. 3b) is at 0.34 $\mu s^{-1}$, just above the experimentally estimated $\sigma$-value above 29 K. This shows that there are many possible muon sites in the unit cell. However, a closer inspection of Fig. 3b reveals small spheres located at the (0.5, 0, 0.25) position, indicating a minimum at a large interstitial site, which leads us to assign a probable muon site in $CeOs_2Al_{10}$. However, this muon site is at variance with the muon site '4a' in $CeRu_2Al_{10}$ determined by Kambe et al [22]. This presumed difference in the muon stopping sites between the two compounds might explain the differences in the μSR spectra observed below 29 K in $CeOs_2Al_{10}$ compared with $CeRu_2Al_{10}$. The spectra below 29 K are best described by three oscillatory terms and an exponential decay, as given by the following equation

$$G_z(t) = \left( \sum_{i=1}^{3} A_i cos(\omega_i t + \varphi) \exp\left(-\frac{(\sigma_i t)^2}{2}\right) \right) + \exp(-\lambda t) + C \quad (4)$$

where $\omega_i = \gamma_\mu H^i_{int}$ are the muon precession frequencies ($H^i_{int}$ are the internal fields at the muon site), $\sigma_i$ are the muon depolarization rates (arising from the distribution of the internal field) and $\varphi$ is the phase. However, one of the terms has been calculated with a very small internal field which is less than one precession frequency. We were therefore unable to determine its precise value and the signal is dominated by the distribution of internal fields, hence it was fixed to zero value.

In Figure 4 (left) we have plotted the internal field (or muon precession frequency) at the muon site as a function of temperature. This shows that the internal fields appear just below 29 K, showing clear evidence for long-range magnetic order. However, the associated internal fields are found to be very small, which either indicates a small ordered state magnetic moment of the $Ce^{3+}$ ion, or the muon site having high spatial symmetry where one expects the cancellation of magnetic field in an antiferromagnetic structure. Given that no magnetic Bragg peaks have been observed in the neutron diffraction data as discussed above we must conclude that the ordered magnetic moment of the $Ce^{3+}$ ion is small, however without a prior knowledge of the magnetic structure we can not estimate the precise value of the magnetic moment. Now examining the temperature dependence of the internal fields, we can see that there is a dip in the internal field (see Fig. 4 top left), which occurs around 15 K. The occurrence of the dip coincides with both a structural distortion observed in the recent electron diffraction study and with the onset of semiconducting behaviour in the resistivity below 15 K [23]. Moreover, below 15 K the first and the second component of the depolarisation rates also increase (Fig. 4 right). In principle this could originate from various phenomena related to a change in distribution of internal fields, but a structural transition is a likely candidate in view of the structural instability reported in this system [14]. It is to be noted that the third depolarization rate associated with zero-frequency also exhibits dramatic changes with temperature (see the inset Fig. 4 bottom-right).

Now we discuss the effect of applied magnetic field on the µSR spectra of $CeOs_2Al_{10}$ at 5 K in the magnetic ordered state and 35 K in the paramagnetic state (see Fig. 5). Our aim was to see how rapidly the musr spectra change with applied field to get some estimate of the internal field at the muon site. At 5 K µSR spectra in zero field reveal a strong suppression of the oscillations due to a broad distribution of internal fields. Figure 5 illustrates the systematic change in the µSR spectra with applied magnetic fields at 5 K and 35 K. It is evident that for a small longitudinal field (20 G) at 5 K the oscillations reappear. This could be due to the decoupling of the low field term. On increasing the field to 100 G, the oscillations disappear but the full instrumental asymmetry is not recovered until an applied field of 500 G. This may suggest that the internal fields at the muon sites are below 100G, which are in agreement with the values presented in Fig.4. At 35 K, the Kubo-Toyabe term is decoupling in a field >20 G and the asymmetry is flat (see Fig. 5).

It is worthwhile to compare the ZF µSR spectra of $CeFe_2Al_{10}$, which exhibits Kondo insulator behaviour [12], with that of $CeOs_2Al_{10}$. For this purpose, we have carried out a zero-field µSR study between 4 K and 40 mK on $CeFe_2Al_{10}$. The ZF µSR spectra of $CeFe_2Al_{10}$ exhibit a Kubo-Toyabe functional form between 4 K and 40 mK, which is very similar to that found for $CeOs_2Al_{10}$ above 29 K, and for $CeRu_2Al_{10}$ above 27 K [22]. Furthermore, we have found that the electronic component of the relaxation rate increases slightly below 1 K in $CeFe_2Al_{10}$. However, we do not see any clear sign of long-range magnetic ordering down to 40 mK in $CeFe_2Al_{10}$. This is in agreement with the high Kondo temperature and mixed valence behaviour observed in $CeFe_2Al_{10}$.

## (3) Inelastic neutron scattering study

With the dramatic changes observed in the µSR spectra and the gap observed in the susceptibility and heat capacity of $CeOs_2Al_{10}$ at temperatures below 29 K, it is of interest to investigate the spin gap value and its temperature dependence using inelastic neutron scattering (INS). Therefore, we briefly report the temperature dependent INS spectra of $CeOs_2Al_{10}$ in this section. A detailed report on the inelastic neutron scattering investigations on $CeT_2Al_{10}$ (T=Fe, Ru and Os) compounds can be found in ref. [21]. Figure 6 displays the inelastic neutron scattering spectra of $CeOs_2Al_{10}$ at various temperatures at Q=1.58 Å$^{-1}$ measured on the MARI spectrometer. There is a clear magnetic excitation centered around 11 meV, which may be compared to the 8 meV excitation found in the equivalent compound CeRu$_2$Al$_{10}$ [20, 21]. The value of the peak position can be taken as a measure of the spin gap energy in these compounds [29]. The spin gap energy of 11 meV is in good agreement with the value determined from the exponential behaviour of the observed magnetic susceptibility and specific heat ($\Delta_{average}$=117 K =10 meV)  [11].

In the following, we discuss the temperature- and Q-dependence of the spin gap excitation. The 11 meV excitation remains nearly invariant from 4.5 K up to 20 K, i.e. across the 15 K feature found in the µSR response of the first component (see Fig.4 top-left). Towards 26 K however, this resonance evidently diminishes and is rather situated at about 8.2 meV before disappearing at still higher temperatures.  At 30 K, the excitation disappears and the response transforms into a broad quasi-elastic or inelastic feature, at the same temperature at which the oscillations in our ZF-µSR spectra disappear (Figs. 6 and 7). Furthermore the disappearance of inelastic response at 30 K is also in agreement with the temperature 29 K at which the electrical resistivity exhibits a sharp drop, however, it is lower than 45 K at which the magnetic susceptibility exhibits a broad peak.

The Q-dependent integrated intensity between 8 and 14 meV at 2 K nearly follows the Ce$^{3+}$ magnetic form factor squared (F$^2$(Q)), although some very weak oscillating feature around $F^2(Q)$ has been observed [21]. These results indicate that the spin-gap excitation can be mainly attributed to a single-ion nature with some weak effect probably arising from a spin-dimer formation [21]. It is to be noted that the single-ion type response is also observed in the inelastic response of the spin gap system $CeRu_4Sb_{12}$, which does not exhibit any long-range magnetic ordering down to 50 mK [29]. On the other hand, the deviation from a single-ion response is observed in the spin gap system $CeFe_4Sb_{12}$, where it was proposed that the intersite interactions between Ce and Fe are playing an important role [30]. Furthermore very recently some of us have made a detailed investigation of a spin-dimer formation in a polycrystalline sample of the Kondo lattice compound $YbAl_3C_3$ which does not order magnetically down to 50 mK [31], which clearly reveals strong modulation of the spin gap energy as well as its intensity with Q, emphasising the role of the intersite interactions. As the spin gaps in $CeOs_2Al_{10}$ and $CeRu_2Al_{10}$ open up below the magnetic ordering temperature one would expect that the spin gap energy and its intensity would be strongly Q-dependent, but this is not the case. Thus it is a puzzle at present and one needs inelastic neutron scattering measurements on a single crystal of $CeOs_2Al_{10}$ and $CeRu_2Al_{10}$ to finally confirm whether or not we have real single ion-type response arising from an isolated spin-dimer formation or whether it is due to an opening of anisotropic energy gap (arising from CEF effect) in the spin wave spectrum below the ordering temperature. In case of strong CEF anisotropy one needs a finite energy to excite the spin waves and hence energy gap occurs in their excitation spectra.

In order to elucidate the role of the single ion interaction in $CeOs_2Al_{10}$, we proceed to compare the spin gap energy of $CeOs_2Al_{10}$ with that of other non-magnetic spin gap systems. The spin gap energies of the latter measured through inelastic neutron scattering exhibit a universal scaling relation with the Kondo energy ($T_K$) derived from the maximum in the susceptibility [32]. According to the single impurity model [32, 33], we can estimate the high temperature Kondo temperature $T_K$ through the maximum $T_{max}(\chi)$ in the bulk susceptibility as $T_K = 3*T_{max}(\chi) = 135$ K (11.6 meV) for $CeOs_2Al_{10}$. Therefore it would be interesting to show a universal relation between the inelastic peak position (or spin gap energy) and the high temperature Kondo temperature $T_K$ or $T_{max}(\chi)$, as shown in ref. [32], for $CeOs_2Al_{10}$ (in this compound 4f-conduction electron hybridization is strong). In Fig. 8 we have plotted the peak position of the inelastic neutron scattering versus $T_K = 3*T_{max}(\chi)$ for many Ce and Yb based intermediate valence and heavy fermion compounds. We observe an excellent universal scaling relation between $T_{max}(\chi)$ or $T_K$ (see refs. [30, 32] for details), and the inelastic peak position (i.e. the spin gap energy) for the $CeOs_2Al_{10}$ compound and also for $CeRu_2Al_{10}$ (see Fig.8).

## IV. Conclusions

We have carried out neutron diffraction, µSR, and inelastic neutron scattering measurements on $CeOs_2Al_{10}$ to understand the unusual phase transition observed in heat capacity, magnetic susceptibility and electric resistivity as well as to test the theoretical predictions of spin-dimer formation based on Hanzawa's model [15]. The neutron diffraction study does not reveal any clear sign of magnetic ordering or structural distortion, from which we are led to conclude an ordered moment value of less than 0.20(±0.05) $\mu_B$ for the Ce ions. A major achievement of this work has been the finding of frequency oscillations in our µSR spectra, convoluted to at least three distinct frequencies, which for the first time establishes the onset of long-range magnetic ordering in this compound. The temperature dependence of the µSR frequencies and muon depolarization rates follow an unusual behavior with further cooling of the sample below 18 K, pointing at the possibility of another phase transition below 15 K. A comparison of µSR spectra of $CeOs_2Al_{10}$ and $CeRu_2Al_{10}$, compounds which reveal very similar phase transitions in the resistivity and heat capacity, indicates that the magnetic ground state of the former is much more complex than the latter: the former exhibits three frequencies in the µSR spectra, while the latter exhibits two frequencies in the µSR spectra. The inelastic neutron study has established the formation of a spin energy-gap with an energy scale of 11 meV. Furthermore, the temperature dependence of the inelastic peak position and one of the µSR frequencies follow a very similar behaviour. Detailed µSR and neutron scattering measurements on a single crystal sample of $CeOs_2Al_{10}$ are essential to understand the true nature of the spin gap.

**Acknowledgement:**

Some of us DTA/ADH/VKA/ would like to thank CMPC-STFC for financial support. The work at Hiroshima University was supported by a Grant-in-Aid for Scientific Research on Innovative Area "Heavy Electrons" (20202004) of MEXT, Japan. AMS thanks the SA-NRF (Grant 2072956) and UJ Research Committee for financial support.

**Figure captions**

Fig.1 (color online) Neutron powder diffraction patterns of $CeOs_2Al_{10}$ from one of the detector banks of the GEM diffractometer at 2 K, 20 K and 35 K. The solid line represents the profile fit using space group Cmcm. The insets compare the data at three temperatures, 2K (black line), 20K (blue line), and 35 K (red line). The x-axis shows the d-spacing and the y-axis shows the intensity. The top, middle and bottom insets are from the detector banks at scattering angle, 9.4, 18 and 35 degree of the GEM diffractometer. Please note that the different d-ranges of the insets is used to emphasise that there is no clear magnetic scattering over the d-range from 1 to 30 Å at 2 and 20K.

Fig. 2 (color online) The zero-field (ZF) μSR spectra at various temperatures of $CeOs_2Al_{10}$. The insets show the behaviour at shorter time scales.

Fig. 3 (color online) (a) The orthorhombic unit-cell of $CeOs_2Al_{10}$. The muon position is indicated by the yellow spheres. Ce, Os, and Al atoms are symbolized by green, blue and black spheres, and (b) Iso-surface plot of the muon nuclear depolarisation rate with possible muon sites with the position coordinate of (0.5, 0, 0.25). The most probable positions are shown by the red surface in the unit cell.

Fig. 4 (color online) Fit parameters of zero-field (ZF) μSR spectra of $CeOs_2Al_{10}$, internal field vs temperature (Left), and depolarization rate vs temperature (Right). The inset shows the depolarization rate for the lowest frequency, which has been fixed to zero value in our analysis.

Fig. 5 (color online) μSR spectra as a function of applied magnetic fields at 5 K (bottom) and 35 K (top).

Fig. 6 (color online) Inelastic neutron scattering spectra of $CeOs_2Al_{10}$ at different temperatures at Q=1.58 Å$^{-1}$ measured with an incident energy of $E_i$=25 meV. The solid lines represent fits. Dotted, dash-dotted and dash-dot-dotted lines represent the components of the fit (see appendix for details). It is to be noted that 30 K data can be fitted with either a quasi-elastic line or an inelastic line.

Fig.7 (color online) The temperature dependence of the INS peak position (left y-axis) and one of the µSR frequencies (right y-axis).

Fig. 8 (color online) The inelastic peak position versus $T_K=3*T_\chi^{max}$ estimated from the maximum in the DC susceptibility for various Ce-based strongly correlated electron systems, see ref. [32] for details.

**Appendix:**

In this appendix we present a more detailed discussion on the analysis of the inelastic neutron scattering spectra presented in Fig. 6. We have used a Gaussian function convoluted with an exponential decay function (GED) (plus a Gaussian to account for the wings on both sides of zero energy transfer) for the elastic line. The width and decay parameters of the elastic line were estimated by fitting monochromatic vanadium data with the same incident energy. For the analysis of the sample elastic line, the width parameters and the exponential decay parameter were kept fixed. We only varied the intensity of GED, while keeping the intensity of the Gaussian (use to account for the wings) bound to that of GED (the binding ratio was obtained from the fit of the vanadium data). A Lorentzian form of the spectral function convoluted with the instrument resolution was used for both inelastic and quasi-elastic lines. The instrument resolution parameters were estimated first by fitting the monochromatic vanadium run mentioned above. The following form of S(Q, ω) was used for the inelastic and quasi-elastic responses:

$$S(Q,\omega) = \left(\frac{\omega}{(1-\exp(-\hbar\omega/k_B T))}\right) F^2(Q) \frac{1}{2} \sum_i (\chi_i) \left(\frac{\Gamma_i}{(\hbar\omega - \hbar\omega_{i0})^2 + \Gamma_i^2} + \frac{\Gamma_i}{(\hbar\omega + \hbar\omega_{i0})^2 + \Gamma_i^2}\right)$$

Where $F^2(Q)$ is the square of the $Ce^{3+}$ form factor taken from ref. [34], $\Gamma_i$ is the linewidth (HWHM) and $\omega_{i0}$ is the position of the peak and $\chi_i$ is the static susceptibility, which is proportional to the integrated intensity of the peak. We have analysed data at all temperatures using one inelastic peak, while at 30 K data we also analysed using a quasi-elastic line. It is to be noted that 30 K data can be fitted either with an inelastic peak (at 6.0±4.0)) or a quasi-elastic peak (one need high resolution data to resolve inelastic and quasi-elastic signals). The quality of the fit to the data can be seen in Fig. 6. The estimated value of the susceptibility at 2 K is $\chi_{INS}$ = 2.12x10$^{-3}$ (±5.62x10$^{-4}$) (emu/mole), which is in good agreement with that measured by dc-magnetization of a single crystal and by generating the polycrystalline average, 3.24x10$^{-3}$ (emu/mole) [11]. The estimated value of the inelastic linewidth at 2 K is 1.52 (±0.5) meV.

Table I Atomic coordinates (x, y, z), isotropic displacement parameters ($U_{iso}$), site occupancies, and selected Ce-Ce interatomic distances (Å) determined from the full structure refinement using GSAS at 2 K for $CeOs_2Al_{10}$. The orthorhombic lattice parameters at 2 K, 20 K and 35 K are given for comparison. The profile R-factor is $R_p$=3.2 % and the weighted profile R-factor is $R_{wp}$=5.8% Estimated standard deviations as obtained by the Rietveld fit are given in parentheses.

| Atom | x | y | z | $U_{iso}$ (Å$^2$) | Occupancy |
|---|---|---|---|---|---|
| Ce | 0 | 0.1257(2) | 0.25 | 0.00111 (4) | 0.981(5) |
| Os | 0.25 | 0.25 | 0 | 0.00003 | 1 |
| Al1 | 0.2240(2) | 0.36530(2) | 0.25 | 0.0036 (6) | 0.988(6) |
| Al2 | 0.3494(2) | 0.1324(2) | 0.25 | 0.0029 (4) | 1 |
| Al3 | 0 | 0.1579(2) | 0.6020(2) | 0.0005 (4) | 1 |
| Al4 | 0 | 0.3779(3) | 0.0485(2) | 0.0002 (5) | 0.987(6) |
| Al5 | 0.2240(2) | 0 | 0 | 0.0020(4) | 1.009(3) |

Lattice parameters and selected Ce-Ce interatomic distances (in Å)

| | 2K | 20K | 35K |
|---|---|---|---|
| a | 9.1216(2) | 9.12180(2) | 9.1221(2) |
| b | 10.2543(2) | 10.2545(2) | 10.2545(2) |
| c | 9.1690(2) | 9.1691(2) | 9.1694(2) |
| Ce-Ce (2) | 5.260(2) | 5.260(2) | 5.260(2) |
| Ce-Ce (4) | 6.862(1) | 6.862(1) | 6.862(1) |
| Ce- Ce (4) | 6.951(1) | 6.951(1) | 6.951(1) |

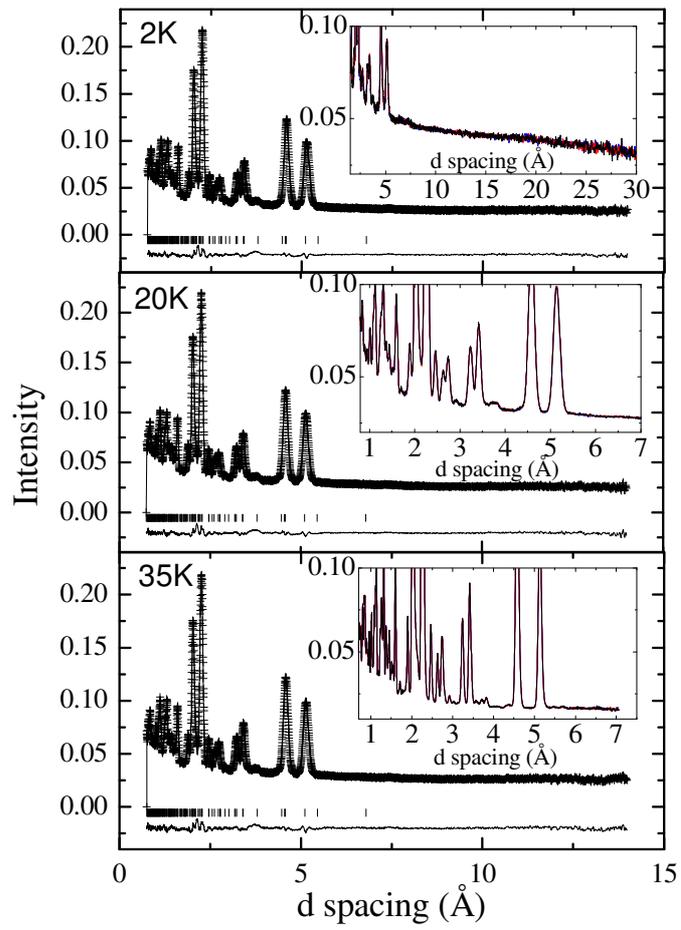

Fig.1 Adroja et al

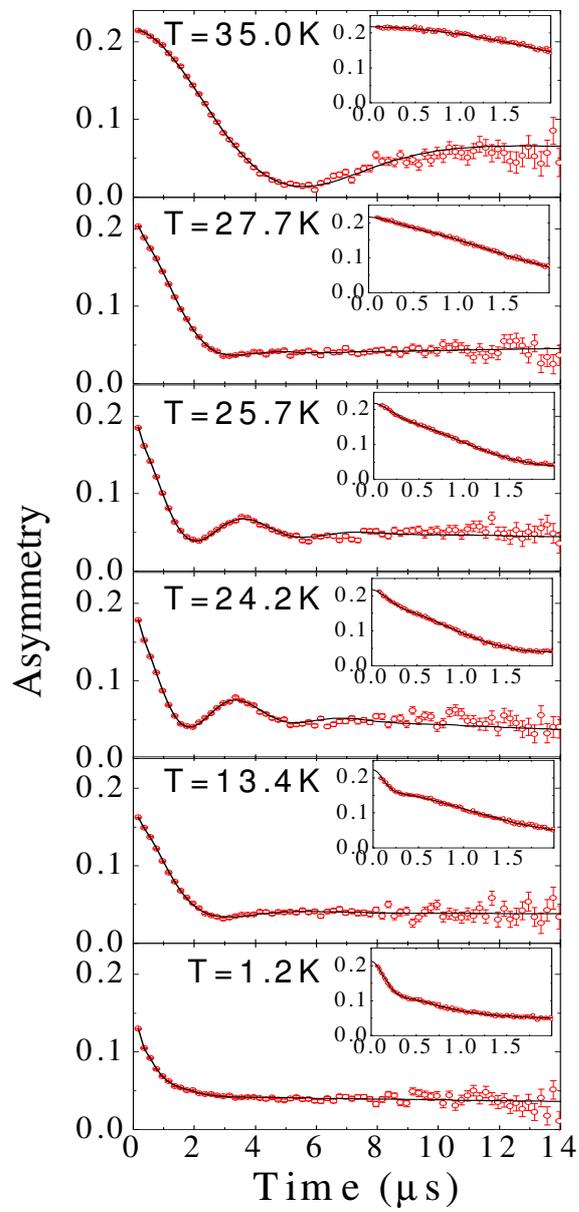

Fig.2 Adroja et al

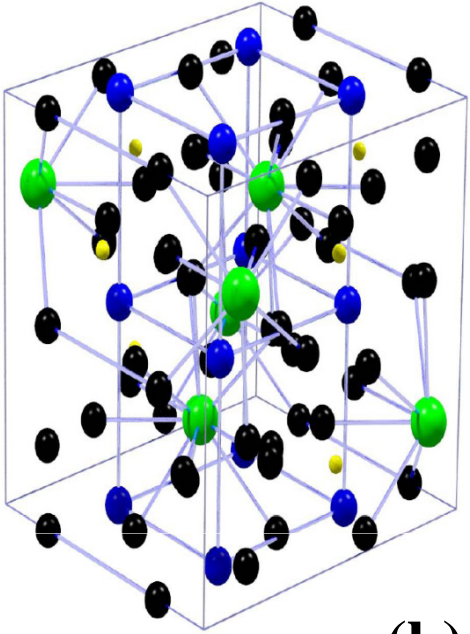

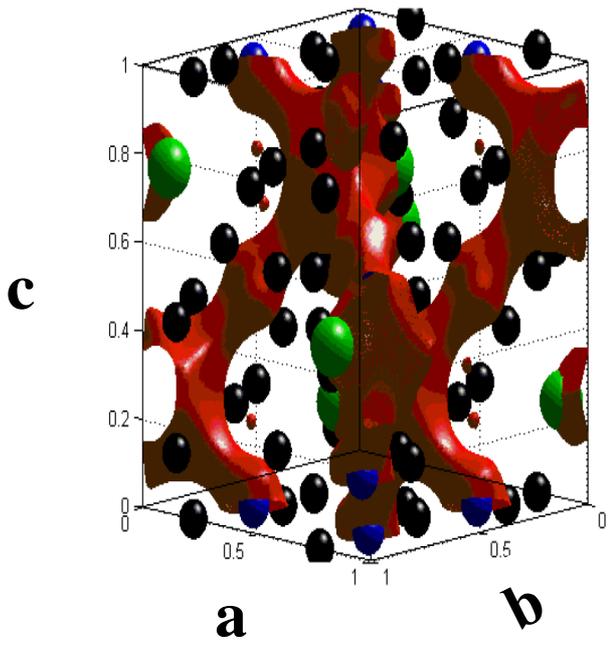

Fig.3 Adroja et al

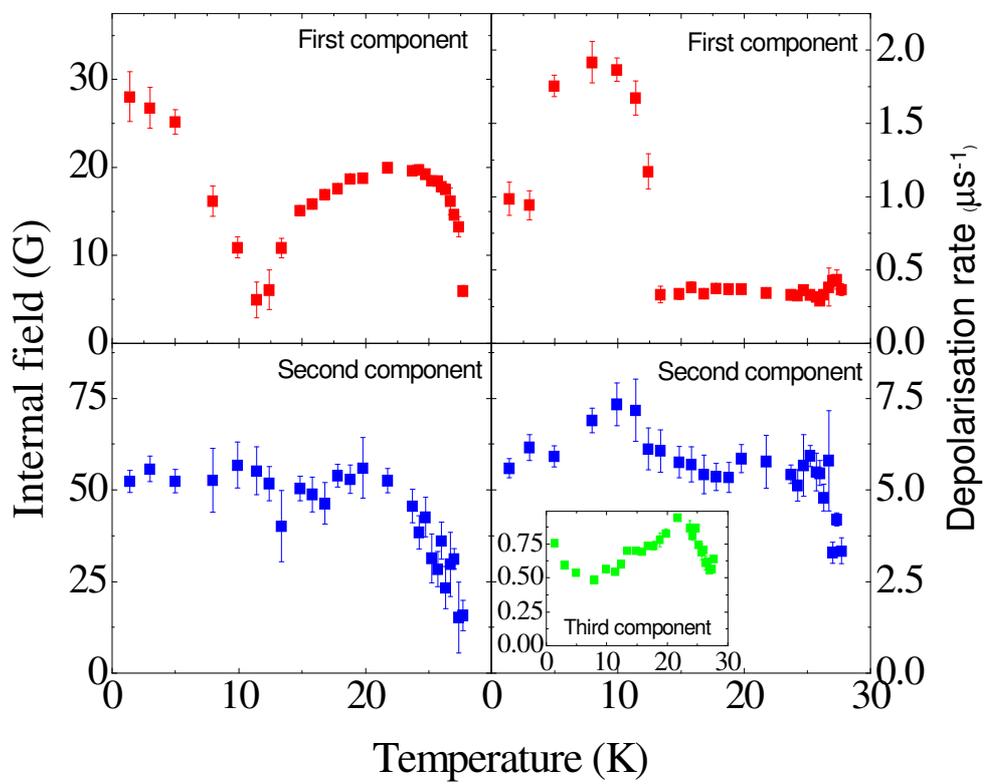

Fig.4 Adroja et al

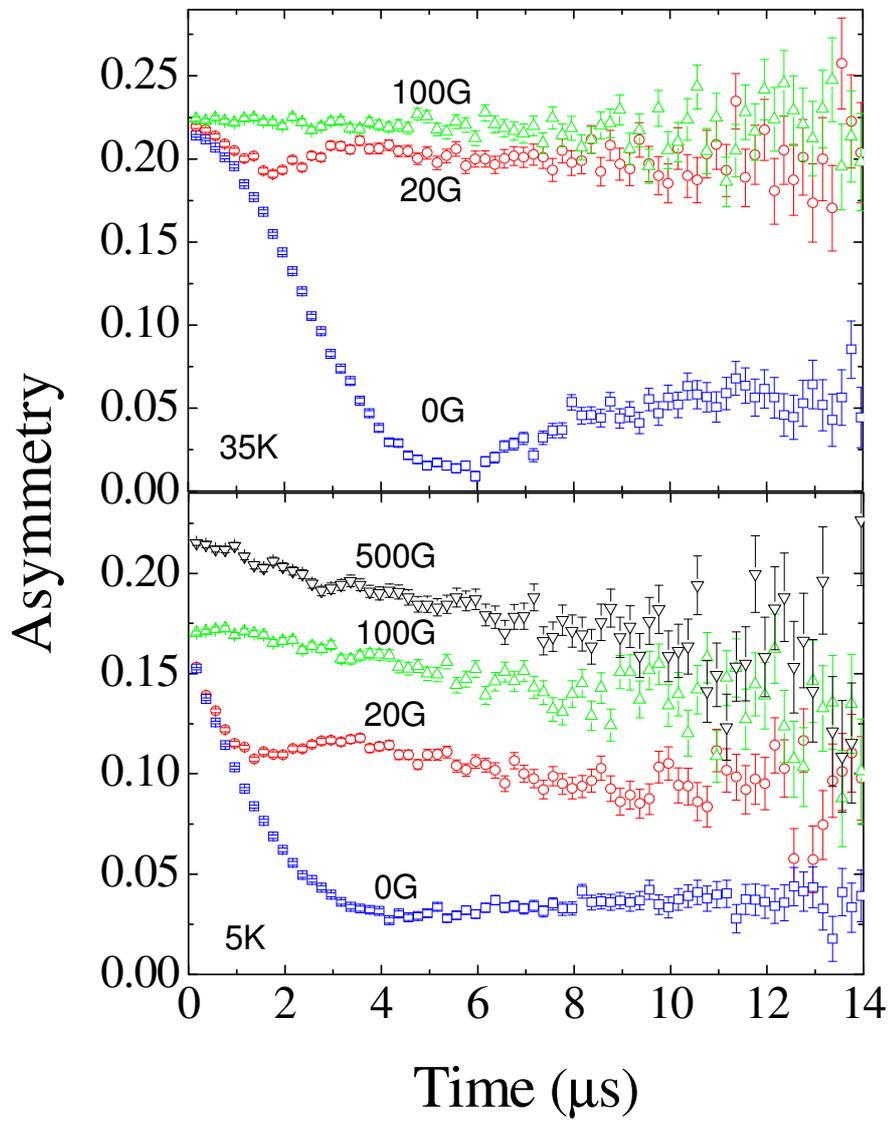

Fig.5 Adroja et al

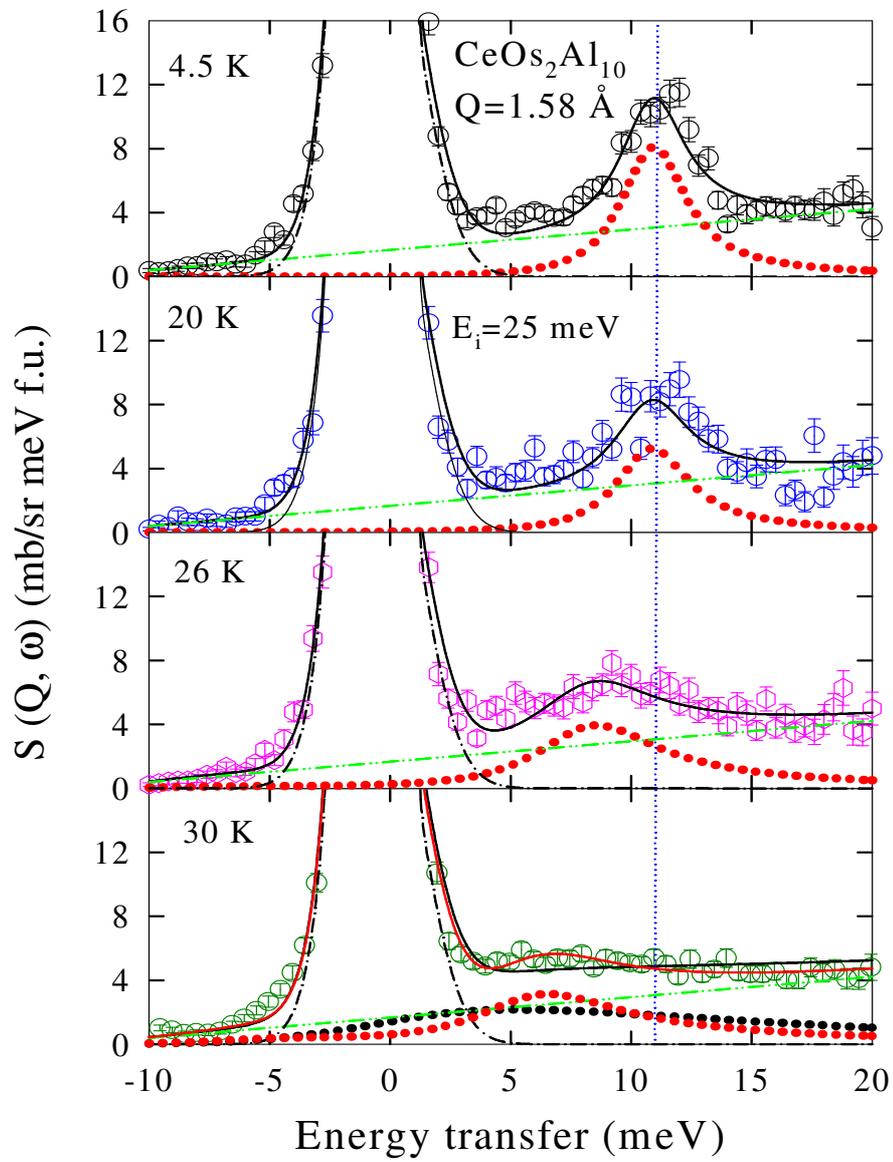

Fig.6 Adroja et al

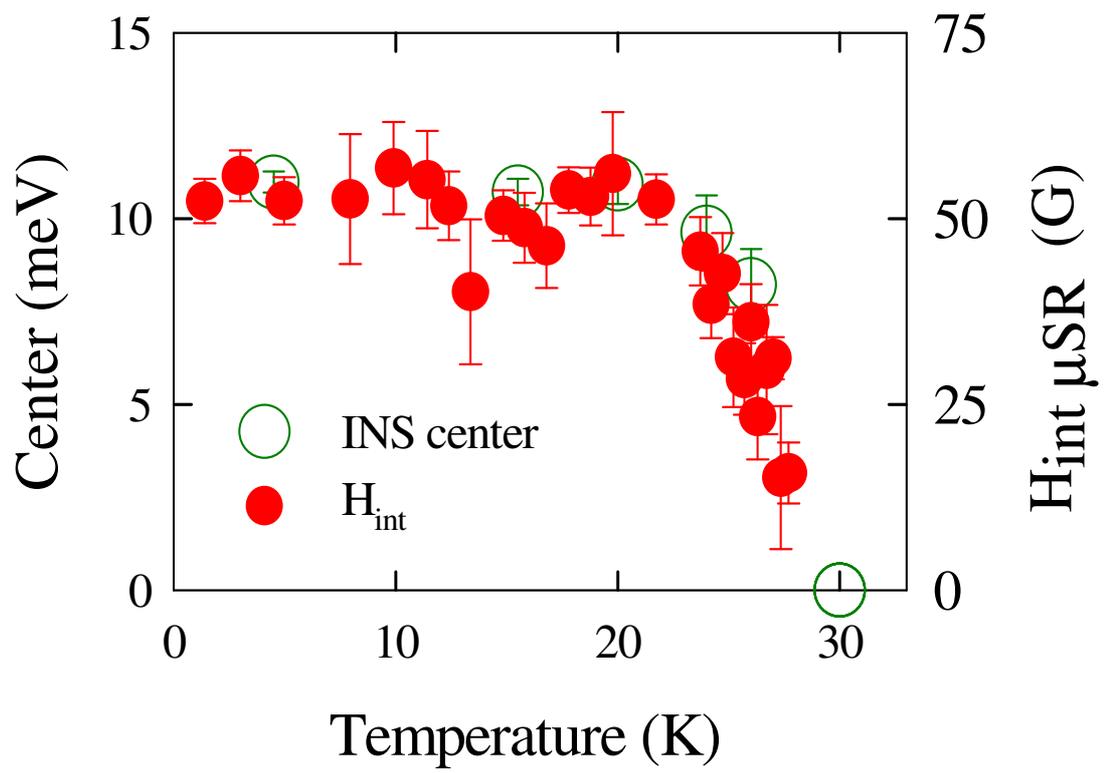

Fig.7 Adroja et al

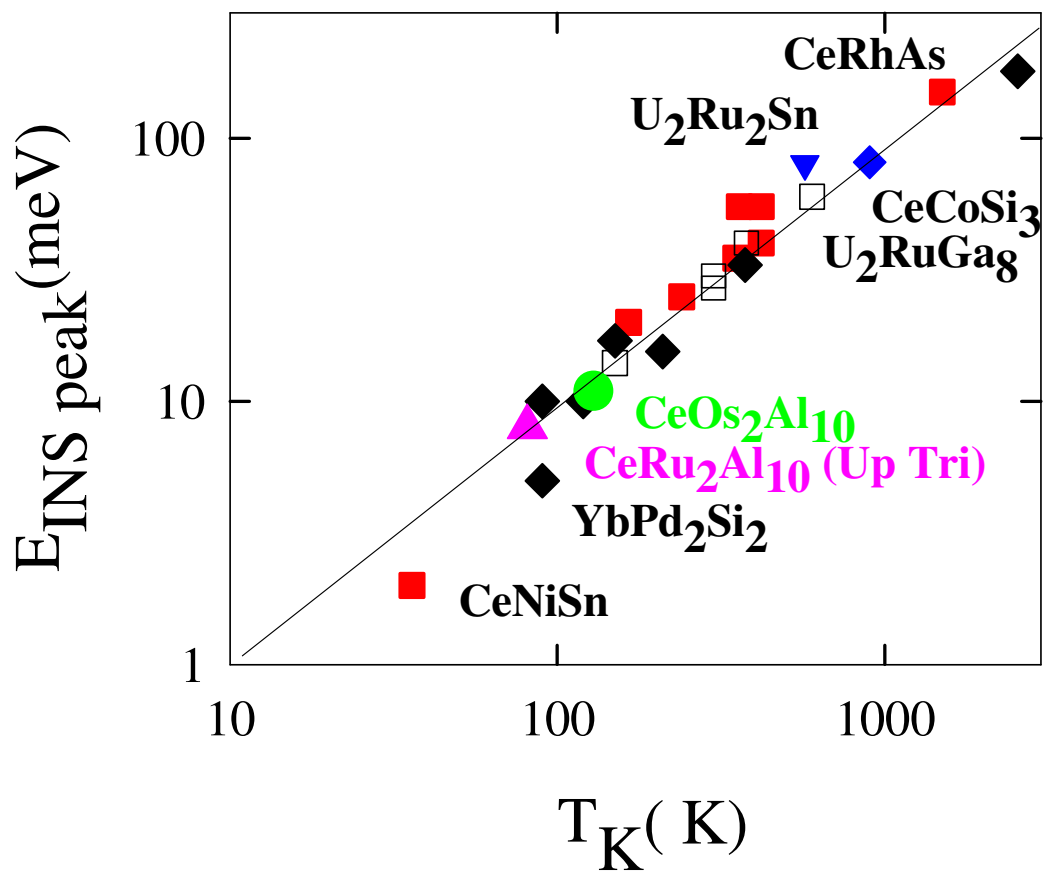

Fig.8 Adroja et al